\documentclass{llncs}

\usepackage{times}
\usepackage{epsfig}
\usepackage{latexsym}
\usepackage{amsmath,amscd,amssymb}
\usepackage{url}
\usepackage{graphicx}
\usepackage{xspace}
\usepackage{paralist}
\usepackage{enumerate}
\newcommand{\leanparagraph}[1]{\smallskip\noindent\textbf{#1}. }

\newcommand\hex{{\sc hex}}
\newcommand\dlvhex{{\sc dlvhex}}

\newcommand{\nop}[1]{}

\title{Integrated Algorithms for HEX-Programs\\ and
Applications in Machine Learning}

\author{Tobias Kaminski}

\authorrunning{Tobias Kaminski}

\institute{Institute of Logic and Computation, TU Wien, Vienna, Austria\\
\email{kaminski@kr.tuwien.ac.at}}

\begin{document}
\maketitle

\begin{abstract}
This paper summarizes my doctoral research on evaluation algorithms for \hex-programs, which extend Answer Set Programming with means for interfacing external computations.
The focus is on integrating different subprocesses of \hex-evaluation, such as
solving and external calls as well as grounding, and on applications of \hex-programs
in the area of Machine Learning.
\end{abstract}

\section{Motivation}
Due to current trends in distributed systems and information integration, there is an increasing need for accessing external information sources from within knowledge representation formalisms such as \emph{Answer Set Programming} (\emph{ASP}) \cite{GelfondL91}. 
For instance, it might be necessary to integrate information derived from a (possibly remote) \emph{Description Logic} (\emph{DL}) ontology into the computation of an answer set. If the derivation in the ontology is relative to information in the ASP part, a bidirectional exchange between a DL reasoner and an ASP solver is required.
This kind of interaction is not provided by ordinary ASP, and pre-computing all possible derivations from the ontology and adding them to the answer set program is often not feasible.
Motivated by this, the \textsc{hex}-formalism \cite{eiter2016model} has been developed, where external sources can be referenced in a program, and are evaluated during solving.
The approach is related to \emph{SMT}, but the focus is more on techniques for evaluating general external sources represented by arbitrary computations, i.e.\ it enables an API-like approach such that a user can define plugins without expert knowledge on solver construction.

By employing \textsc{hex}, a user can, e.g., define a library function for concatenating strings, accessed via an external predicate $\&concat$.
It could be used as illustrated by the following rule, where a first name and a last name are provided, and $\&concat$ returns the full name:
$ \textit{fullname}(\textit{Full}) \leftarrow \&\textit{concat}[\textit{F},\textit{L}](\textit{Full}), \textit{firstname}(\textit{F}), \textit{lastname}(\textit{L}).$

\textsc{hex}-programs are very expressive since they enable a bidirectional exchange of information between a logic program and external sources and thus, encompasses the formalization of nonmonotonic and recursive aggregates. Consequently, \textsc{hex} is suited for a wide range of applications, but also requires sophisticated evaluation algorithms to deal with the complexity that goes along with the high expressiveness. For this reason, my thesis work aims at the design and implementation of novel integrated solving techniques for improving the
efficiency of the formalism in general, as well as for specific classes
of programs.
Another focus of my work is on new applications that leverage the provided techniques, and in turn push the advancement of the formalism.

\section{Goals of My PhD Thesis}
Challenges regarding efficient evaluation of \textsc{hex}-programs comprise the lack of a tight integration of the solving process with the evaluation of external sources and with the grounding procedure.
Accordingly, the main goals of my doctoral research are:
\begin{enumerate}
\item to design advanced reasoning techniques that improve the evaluation of \textsc{hex}-pro\-grams by tightly integrating processes which have so far been treated as mostly independent sub-problems.
\item to develop innovative applications of \textsc{hex}-programs
that utilize external atoms for integrating as well as realizing methods from the area of \emph{Machine Learning} (\emph{ML}).
\item to implement newly developed evaluation algorithms in the \textsc{hex}-program solver \textsc{dlvhex} \cite{EiterIST06}, and to investigate their performance using benchmark problems.
\end{enumerate}

\section{Background}
Here, I start by briefly summarizing the work most related to \hex\ and its applications, and I  introduce the theoretical background which my thesis work is based on.

\leanparagraph{Related Work}
As there are many scenarios where it is more natural, and often more efficient, to outsource some information or computation in ASP, several approaches exist for this purpose, realizing different degrees of integration.
\textsc{dlv-ex} programs \cite{CalimeriCI07} represent an early approach, which enables bidirectional communication with an external source, and allows the introduction of new terms by \emph{value invention}. 
The \textsc{clingo} system also provides a mechanism for importing the extension of user-defined predicates via function calls during grounding \cite{GebserKKS14}.
In both cases, the interaction is more restricted than in \hex{} such that, e.g., nonmonotonic aggregates cannot be expressed.

\textsc{clingo 5} \cite{GebserKKOSW16} provides generic interfaces for combining theory solving with ASP, but its semantics differs from \hex\ and the approach is targeted at system developers.
Besides, there are extensions of ASP towards
the integration of specific sources; e.g.,
the \textsc{clingcon} system \cite{OstrowskiS12} implements \emph{constraint ASP} relying on a tailored integration of a constraint solver.
The setting of \hex{} differs as its goal is to enable a broad range of users to implement custom external sources and to harness efficient solving techniques.

\textsc{hex} has been applied to a wide range of use cases. Among them are a framework for executing scheduled actions in external environments (\emph{Act}\textsc{hex} \cite{FinkGIRS13}); a system for merging belief sets based on nested \textsc{hex} programs (\emph{MELD system}, \cite{RedlEK11}); and an artificial agent able of playing the computer game \emph{Angry Birds} (\emph{Angry}\textsc{hex}, \cite{EiterFS16}).

\leanparagraph{\textsc{hex}-Programs}
\textsc{hex}-programs \cite{eiter2016model} extend ASP by allowing the use of \emph{external atoms} of the form $\&g[p_1,...,p_k](c_1,...,c_l)$ in rule bodies, where $\&g$ is an \emph{external predicate} name, $p_1,...,p_k$ are input predicate names or constants, and $c_1,...,c_l$ are output constants. The ground
semantics of an external atom $\&g[p_1,...,p_k](c_1,...,c_l)$ is given by a \emph{Boolean} $1+k+l$-ary oracle function $f_{\&g}$ s.t.\ an external atom evaluates to \emph{true} for a given assignment $\mathbf{A}$ over ordinary atoms if the oracle function returns \emph{true}, i.e.\ $f_{\&g}(\mathbf{A},p_1,...,p_k,c_1,...,c_l) = \mathbf{t}$, and to \emph{false} otherwise. The notion of satisfaction is extended to \textsc{hex}-rules and programs in the obvious way. Answer sets of a \textsc{hex}-program~$\Pi$ are those assignments $\mathbf{A}$ to ordinary atoms which are minimal models of the program consisting of all rules in $\Pi$ of which the body is satisfied under $\mathbf{A}$ (the so-called \emph{FLP-reduct} \cite{FaberLP04}, an alternative to the well-known \emph{GL-reduct}).

The basic procedure for computing the answer sets of a \textsc{hex}-program $\Pi$ consists in replacing each (ground) external atom $\&g[p_1,...,p_k](c_1,...,c_l)$ by an ordinary atom of the form $e_{\&g[p_1,...,p_k]}(c_1,...,c_l)$ and adding a guess $e_{\&g[p_1,...,p_k]}(c_1,...,c_l) \vee ne_{\&g[p_1,...,p_k]}(c_1,...,c_l) \leftarrow$ for its evaluation \cite{TUW-140622}. The result of this translation is an ordinary answer set program; and ordinary ASP solvers such as \textsc{clasp} can be used for computing \emph{model candidates}.
However, guesses for external atoms must be checked afterwards for compatibility with the external semantics.
By integrating \emph{Conflict-Driven Nogood Learning} (\emph{CDNL}) search into the \textsc{hex}-algorithm, the input-output relations can be learned from these checks in form of \emph{nogoods} to avoid wrong guesses in the future search.
Moreover, even if a model candidate complies with the answers of the corresponding oracle calls, it still needs to be checked for minimality wrt.\ the FLP-reduct.

\section{Research Progress}
In this section, I present the research results obtained since I started my PhD studies.

\subsection{Integration of Solving and External Evaluations}
In the beginning of my doctoral research, I worked on the tighter integration
of the solving process and external calls \cite{DBLP:journals/jair/EiterKRW18}, which required an extension
of the oracle semantics.
Before, oracle functions were only defined for complete inputs to external atoms, such that they could only be evaluated after the whole input was decided.
As a result, many wrong guesses could only be detected late during search and nogoods were large as they usually entailed the complete input assignment.
However, this could not be improved when using two-valued assignments since external sources might be nonmonotonic, and they are \emph{black boxes} such that theory specific techniques like in SMT cannot be applied.

We have overcome the mentioned challenges by extending the two-valued semantics to three-valued assignments that use the classical values \emph{true} and \emph{false}, and the new value \emph{unassigned}.
Based on partial assignments, we have introduced new evaluation techniques to increase the performance of \textsc{hex}-evaluation, which can be utilized in the search for model candidates as well as the search applied during checking minimality wrt.\ the FLP-reduct.
First, we have extended two-valued oracle functions to three-valued ones, which allows evaluation at any point during search under partial input.
Second, nogoods now can also be learned under partial assignments, which are often significantly smaller.
Moreover, given some input-output nogood, we obtain a set of minimal nogoods by applying a minimization procedure similar to the one in \cite{OstrowskiS12}.
As an alternative, we also incorporated the \textsc{QuickXplain} algorithm \cite{Junker04} for conflict minimization, which is more suited for nogoods that contain many irrelevant literals.
The benefit of the new solving techniques has been verified by experiments using \dlvhex.

\subsection{Integration of External Sources and Minimality Checking}
In addition to the usual minimality check of ASP,
a special minimality check wrt.\ the FLP-reduct is required during
the evaluation of \hex-programs
to avoid
cyclic justifications via external sources.
The check is a bottleneck in practice
as it often accounts for most of the time required to evaluate \hex-programs.
For this reason,
syntactic information regarding atom dependencies has been used to detect
situations where 
the external minimality 
check can be skipped \cite{efkrs2014-jair}.
However, this approach overapproximates the real dependencies
as a result of the black-box nature of external sources.

In our most recent work \cite{DBLP:conf/lpnmr/EKLPNMR2019}, we considered a tighter integration of minimality checking and external sources by showing how the real external dependencies can be approximated more closely
by also taking semantic dependency information into account. 
The additional dependency information can be provided by a user or even generated automatically. 
This brings us closer to a clear-box view of external sources,
and allows us to skip the external minimality check in more cases.
Furthermore, we 
stated conditions
under which the costs for checking and generating
semantic dependency information can be reduced.
Using an experimental evaluation, we could verify that having more fine-grained information about the actual dependencies among atoms is crucial for applications where otherwise the overestimation makes the minimality check infeasible.

\subsection{Integration of Grounding and Solving}
During the second year of my PhD studies, I worked on integrating
the lazy-grounding ASP solver \emph{Alpha} \cite{DBLP:conf/lpnmr/Weinzierl17} into the
\dlvhex{} system, with the goal to achieve a tighter integration of \hex-sovling and grounding.
Lazy grounding avoids an exponential blowup of the grounding by interleaving
grounding and solving, whereby rules are grounded on-the-fly depending on the satisfaction
of their bodies.
The resulting approach exhibits promising results for classes of programs
where the grounding bottleneck of ASP is an issue \cite{DBLP:conf/ijcai/EiterKW17}.
This issue is even more challenging to tackle within the framework of \hex\
due to the need for grounding external atoms; and 
nonmonotonic dependencies and value
invention (i.e., import of new constants) from external sources make
the integration nontrivial.
As a result, we needed to introduce a novel external source interface to 
incrementally extend a \hex-program grounding, where new output
terms may appear \emph{during solving}. This resulted in a
novel evaluation algorithm for \hex-programs that
can incorporate lazy-grounding solvers as backend solvers.

\subsection{Applications of HEX-Programs in Machine Learning}

As \hex\ allows to integrate different formalisms,
it is well-suited for combining diverse forms of reasoning.
In this branch of my research, my goal was to exploit this strength for two new applications in the area of ML.
The first one integrates an external statistical classifier,
while the second encodes an existing approach for logic-based ML.

\subsubsection{Hybrid Classification of Visual Objects}
A basic task in \emph{Statistical Relational Learning} \cite{getoor2007introduction} is \emph{Collective Classification} \cite{SenNBG10},
which
is simultaneously finding
correct labels for a number of interrelated objects, e.g., predicting the classes of objects in a
complex visual scene.
Even if advanced algorithms for object recognition have
been developed, 
they may fail unavoidably and yield ambiguous results due to few
training data, noisy inputs, or inherent ambiguity 
of visual appearance.  
It is then still possible to
draw on further information from the context in which
an object occurs to disambiguate its label. 

We approached the Collective Classification problem in ASP by defining \emph{Hybrid Classifiers} (\emph{HC}) that combine a local classifier, which 
predicts the probability of each local label 
based on object features, with context constraints (weighted ASP constraints) using object relations \cite{EiterK16}.
At this, external atoms of \hex\ can be used to interface an ontology reasoner, a spatial reasoning calculus as well as the local classifier directly from within the encoding.
This has not been realized in the first version of the approach, where the integration was created ad-hoc using a pre-processing step. However, the usage of external atoms for this purpose will be described in my dissertation.

To obtain a probabilistic semantics, we embedded our encoding into the formalism $LP^{MLN}$ \cite{LeeW16}, such that an HC corresponds to an $LP^{MLN}$ program.
We showed that solutions of the resulting \emph{HC encoding} can be
obtained efficiently via a backtranslation from $LP^{MLN}$ into classical ASP with weak constraints \cite{BuccafurriLR00}, and by leveraging combinatorial optimization capabilities of ASP solvers.
Experiments wrt.\ object classification in indoor and outdoor scenes exhibit significant accuracy improvements compared to using only a local classifier.

\subsubsection{Meta-Interpretive Learning}
In the area of \emph{Inductive Logic Programming} (\emph{ILP}),
\emph{Meta-Interpretive Learning} (\emph{MIL}) is a recent approach, introduced by Muggleton et al.\ \cite{MuggletonLT15}, that learns logic programs from examples and background knowledge
by instantiating meta-rules.
The Metagol system \cite{metagol} efficiently
solves MIL-problems by relying on the query-driven search of
Prolog.
Its focus on positive examples, however, effects that Metagol
can detect the derivability of negative examples only at a later check,
which can severely hit performance.
Viewing MIL-problems as
combinatorial search problems, they can alternatively be solved by
employing ASP, which may result in performance
gains as a result of efficient conflict propagation.
By employing modern ASP solvers, violations of
negative examples can potentially be propagated earlier.

However, a
straightforward ASP-encoding of MIL results in a huge search space due
to a lack of procedural bias and the need for grounding. To address
this challenge, we have encoded MIL in the \hex-formalism \cite{DBLP:journals/tplp/KaminskiEI18}.
Our encoding utilizes external atoms to outsource background knowledge that defines manipulations of complex
terms such as lists and strings, which is easy to realize in Prolog but less supported by ASP.
Moreover, we identified a class of MIL-problems which can be solved efficiently by using our \hex-encodings, and showed empirically that the performance can be increased compared to Metagol by employing \hex.
In addition, by abstracting from term manipulations in the encoding
and by exploiting the \hex-interface mechanism, the import of
constants can be entirely avoided in order to mitigate the grounding
bottleneck.

\section{Future Work}
The overarching theme of my thesis is to tightly integrate different mechanisms employed during
\hex-solving,
and a number of new evaluation techniques has been developed for this purpose.
However, there are several further ways in which these techniques can be
combined, extended and exploited for different parts of solving in the future.

First, while we have only employed simple heuristics for deciding the frequency
of external evaluations during \hex-solving, dynamic heuristics could also be used,
where the frequency is adjusted according to the amount of information gained from previous 
calls.
Second, our \hex-algorithm that exploits lazy-grounding could be combined with a pre-grounding
algorithm, where the respective grounding mechanisms are applied for different modules of a \hex-program
based on their properties.
Moreover, additional semantic dependency information, which we used for deciding the necessity of the external minimality check, is also valuable for reducing the number of external evaluations during the model search and grounding, and could be utilized there as well.

\bibliographystyle{plain}
\bibliography{lpnmr_abstract}
\end{document}